\documentclass[10pt,aps,prl,twocolumn,groupedaddress,showkeys]{revtex4-1}
\usepackage{amsmath}
\usepackage{graphicx}
\DeclareGraphicsExtensions{.pdf,.eps,.png,.jpg,.mps}
\usepackage{epstopdf}
\usepackage{threeparttable}
\usepackage{color}

\begin{document}
\title{Strongly coupled coherent phonons in single-layer MoS$_2$}
\author{C. Trovatello$^1$, H. P. C. Miranda$^2$, A. Molina-S\'{a}nchez$^3$, R. Borrego Varillas$^1$, C. Manzoni$^4$, L. Moretti$^1$, L. Ganzer$^1$, M. Maiuri$^1$, J. Wang$^5$, D. Dumcenco$^6$, A. Kis$^6$, L. Wirtz$^7$, A. Marini$^8$, G. Soavi$^{5,9}$, A. C. Ferrari$^5$, G. Cerullo$^{1,4}$, D. Sangalli$^8$, S. Dal Conte$^1$}
\affiliation{$^1$Dipartimento di Fisica, Politecnico di Milano, Piazza L. da Vinci 32, I-20133 Milano, Italy}
\affiliation{$^2$Institute of Condensed Matter and Nanoscience (IMCN), Universit\'e catholique de Louvain, B-1348 Louvain-laneuve, Belgium}
\affiliation{$^3$Institute of Materials Science (ICMUV), University of Valencia, Catedr\'{a}tico Beltr\'{a}n 2, E-46980, Valencia, Spain}
\affiliation{$^4$IFN-CNR, Piazza L. da Vinci 32, I-20133 Milano, Italy}
\affiliation{$^5$Cambridge Graphene Centre, University of Cambridge, 9 JJ Thomson Avenue, Cambridge CB3 0FA, UK}
\affiliation{$^6$Electrical Engineering Institute, Ecole Polytechnique Federale de Lausanne, Lausanne CH-1015, Switzerland}
\affiliation{$^7$Universit\'e du Luxembourg, 162 A, avenue de la Faencerie, L-1511 Luxembourg}
\affiliation{$^8$CNR-ISM, Division of Ultrafast Process in Materials (FLASHit), Area della Ricerca di Roma 1, Monterotondo Scalo, Italy}
\affiliation{$^9$Institut f\"{u}r Festk\"{o}rperphysik, Friedrich-Schiller-Universitat, Jena, Germany}
\begin{abstract}
We present a transient absorption setup combining broadband detection over the visible-UV range with high temporal resolution ($\sim$20fs) which is ideally suited to trigger and detect vibrational coherences in different classes of materials. We generate and detect coherent phonons (CPs) in single layer (1L) MoS$_2$, as a representative semiconducting 1L-transition metal dichalcogenide (TMD), where the confined dynamical interaction between excitons and phonons is unexplored. The coherent oscillatory motion of the out-of-plane $A'_{1}$ phonons, triggered by the ultrashort laser pulses, dynamically modulates the excitonic resonances on a timescale of few tens fs. We observe an enhancement by almost two orders of magnitude of the CP amplitude when detected in resonance with the C exciton peak, combined with a resonant enhancement of CP generation efficiency. \textit{Ab initio} calculations of the change in 1L-MoS$_2$ band structure induced by the $A'_{1}$ phonon displacement confirm a strong coupling with the C exciton. The resonant behavior of the CP amplitude follows the same spectral profile of the calculated Raman susceptibility tensor. This demonstrates that CP excitation in 1L-MoS$_2$ can be described as a Raman-like scattering process. These results explain the CP generation process in 1L-TMDs, paving the way for coherent all-optical control of excitons in layered materials in the THz frequency range.
\end{abstract}
\maketitle

Coherent modulation of the optical properties of a material, following impulsive photo-excitation of the lattice, is fundamentally interesting and technologically relevant because it can be used for applications in sensors\cite{Li2013}, actuators and transducers\cite{Lanzillotti2018,Baldini2018,Baldini2019}, that can be operated at extremely high frequencies (up to several THz\cite{Hase2013}). In order to exploit this effect, it is necessary to understand the mechanism underlying the coherent phonon (CP) generation process and to identify the physical parameters (such as the pump pulse photon energy) that allow their efficient excitation. In view of possible device applications, it is of paramount importance to detect the spectral dependence of the CP amplitude, in order to identify in which photon energy window the optical response of the material can be efficiently modulated.

We present a novel transient absorption (TA) setup, combining broadband detection from 1.8 to 3eV, with extremely high temporal resolution ($\sim$20fs). This is ideally suited to trigger and detect vibrational coherences in different classes of materials. We use it to generate and detect CPs in 1L-MoS$_2$, as a representative semiconducting 1L-transition metal dichalcogenide (TMD). We focus on TMDs because they support strongly bound excitons with unique physical properties, enabling novel applications in optoelectronics and photonics\cite{Koppens2014,Mak2016,Ferrari2015}. However, in TMDs, the dynamical interaction between excitons and phonons, when constrained to 1L, is unexplored.

When TMDs are exfoliated down to 1L, they undergo a transition from indirect to direct band gap\cite{Lebegue2009}, accompanied by a strong enhancement of the photoluminescence (PL) quantum yield (i.e. the ratio between the number of emitted and absorbed photons)\cite{Mak2010}. The thinness of 1L-TMDs leads to a strong reduction of Coulomb screening between electron-hole (e-h) pairs\cite{Chernikov2014}, resulting in enhanced excitonic effects\cite{Molina-Sanchez2013}. The lower energy part of the visible spectrum of all 1L-TMDs is dominated by two excitonic peaks\cite{Mak2010}, commonly referred as A and B, associated to interband optical transitions between the spin-orbit split valence band (VB) and the conduction band (CB) at $K$. At higher energy, interband transitions in the band-nested region between K and $\Gamma$, where VB and CB have the same dispersion, give rise to an additional broad peak: the C exciton\cite{Kozawa2014}. The A/B excitons wavefunctions share the same symmetry of the transition metal $d_{z^{2}}$ orbital in the lowest conduction band state in $K$\cite{Qiu2013}. The C exciton wavefunction has a mixed character and it involves the contribution of the $p_{x}\pm ip_{y}$ chalcogen orbital\cite{Qiu2013,Zhu2011}.

Exciton-phonon coupling in 1L-TMDs has been theoretically and experimentally investigated\cite{Thilagam2016,Slobodeniuk2016,Selig2016,Selig2018,Raja2018,Niehues2018} since it has a large impact on their emission and absorption properties. Light emission in TMD based optoelectronic devices is quenched by exciton-phonon scattering, reducing the number of bright excitons within the radiative light cone\cite{Robert2016}, and leading to the formation of momentum forbidden dark excitons\cite{Selig2018}.

Continuous wave (CW) optical spectroscopies, like PL and absorption, revealed signatures of exciton-phonon interaction in 1L-TMDs\cite{Wang2015,Jones2016,Chow2017,Shree2018}. This gives rise to peculiar processes, such as the appearance of phonon replicas in PL\cite{Wang2015}, excitonic PL upconversion caused by the scattering from charged to neutral excitons\cite{Jones2016} and oscillations of the neutral exciton PL intensity with an energy period matching longitudinal acoustic phonons\cite{Chow2017,Shree2018}. Scattering with phonons contributes to the exciton dephasing process, leading to temperature-dependent broadening of the exciton homogeneous linewidth, as shown by coherent nonlinear spectroscopy\cite{Moody2015}, and determines the linear absorption spectral width\cite{Molina-Sanchez2016}.

Broadband TA spectroscopy probes exciton-phonon scattering in the time domain\cite{Forst2007}. One or more CP modes can be launched by impulsive laser excitation (pump pulse), while the periodical variation of the TA spectrum due to the coherent displacement of the atoms is detected by a delayed broadband probe pulse covering the excitonic transition\cite{Cerullo2007}. This gives direct access to the phase of the coherent lattice displacement and to the phonon dephasing processes caused by lattice anharmonicity\cite{Hase2015}. Another advantage of this technique is the possibility to separately study the CP generation and detection processes\cite{Dekorsy2000,Ishioka2010}. Ref.\citenum{Jeong2016} reported the observation of CP in 1L-WSe$_2$. However, this study was restricted to a narrow energy range ($\sim$100meV) around the A exciton, making it difficult to gain information on how CP excitation affects the whole band structure.

Here, we perform TA measurements on 1L-MoS$_2$ with$\sim$20fs temporal resolution, on a broad photon energy range covering the main excitonic transitions in the visible-UV between 1.8 and 3eV. We find that the C exciton is strongly modulated by the out-of-plane $A'_{1}$ phonon, with an amplitude almost two orders of magnitude larger than previously reported for the A exciton of 1L-WSe$_2$\cite{Jeong2016}. Line shape analysis of the exciton peak dynamics shows that the periodic modulation affects the intensity and the energy of the exciton peak. TA experiments with frequency tunable excitation show that the pump pulse efficiently excites $A'_{1}$ modes when it is tuned around the C exciton. This demonstrates that CP generation and detection are governed by the same Raman-like physical mechanism. \textit{Ab initio} calculations are in good agreement with experiments, and show that CP generation in 1L-MoS$_2$ has the same spectral dependence of the Raman tensor of the $A'_{1}$ mode. Our results show that exciton-phonon coupling is responsible of the strong coherent modulation of the optical properties of TMDs and indicate that these materials are an ideal platform to study the interaction between strongly bound excitons and phonons in the time domain. They also pave the way to coherent all-optical control of excitons in layered materials at THz frequencies.
\begin{figure}
\centerline{\includegraphics[width=90mm]{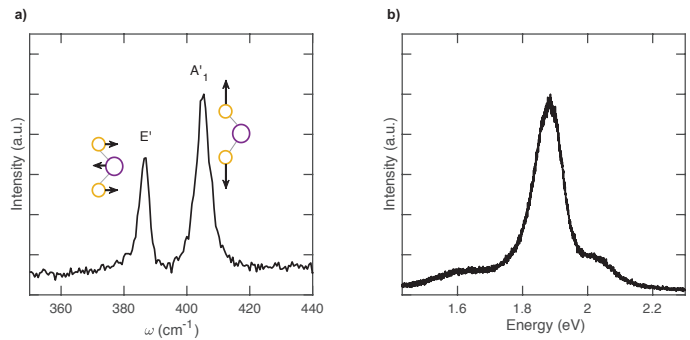}}
\caption{(a) 514nm Raman spectrum of 1L-MoS$_2$. A representation of the atomic displacement along the eigenmodes is also shown. (b) 514nm PL spectrum of 1L-MoS$_2$}
\label{fig:optics-eq}
\end{figure}
\begin{figure*}
\centerline{\includegraphics[width=170mm]{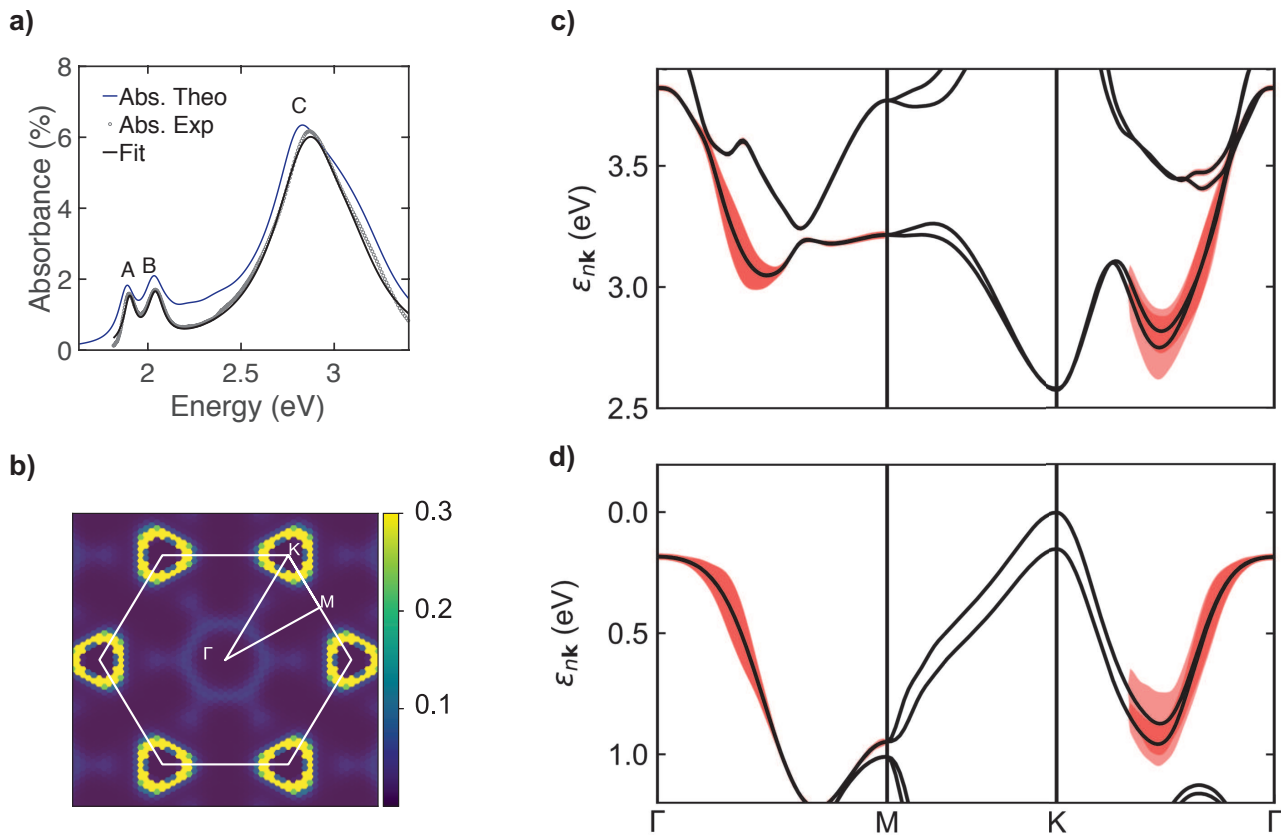}}
\caption{(a) Experimental (grey) and calculated (blue) static absorbance of 1L-MoS$_2$. The black line is a fit with a sum of four Lorentzians. (b) C exciton wavefunction in the the hexagonal BZ (c,d) Calculated quasi-particle band structure of 1L-MoS$_2$. The electronic states participating to the C exciton are in red.}
\label{fig:Simulation-eq}
\end{figure*}

We use large area (i.e. mm sized) chemical vapor deposition (CVD) 1L-TMDs because the minimum spot size of the pump and probe pulses in our high temporal resolution setup is$\sim$150$\mu m$ (see Methods), larger than the typical size of mechanically exfoliated 1L-TMDs. The laser pulses are in fact focused by a spherical mirror, instead of a thick lens or an objective, in order to preserve the high temporal resolution, but at the expense of spatial resolution. 1L-MoS$_2$ is grown on a c-plane sapphire substrate as for Ref.\citenum{Dumcenco2015}. It is then placed on a 200$\mu m$ fused silica (FS) substrate using wet transfer\cite{Dumcenco2015,Bonaccorso2012,Bae2010}. As-grown 1L-MoS$_2$ on sapphire is spin-coated with$\sim$100nm poly(methylmethacrylate) (PMMA). The sample is detached from sapphire in a 30$\%$ potassium hydroxide solution, washed in deionized water, and transferred onto the FS substrate. The PMMA is then dissolved in acetone. The small thickness of the FS substrate prevents possible coherent artifact signals\cite{BritoCruz1988} at zero delay time between pump and probe pulses.

After transfer, the sample is characterized by Raman and PL spectroscopy. For both Raman and PL we use a Renishaw InVia spectrometer with 100x objective at 514nm excitation. The laser power is kept below 100$\mu$W to avoid any possible damage. The Raman spectrum of 1L-MoS$_2$ on FS is reported in Fig.\ref{fig:optics-eq}a. The peak at$\sim$386cm$^{-1}$ corresponds to the in-plane phonon mode ($E'$)\cite{Verble1970,Wieting1971}, while that at$\sim$405cm$^{-1}$ is the out of plane $A'_{1}$\cite{Verble1970,Wieting1971}. The difference between $E'$ and $A'_{1}$ can be used to monitor the number of layers\cite{Lee2010,Li2012}. We obtain$\sim$19cm$^{-1}$, consistent with 1L-MoS$_2$\cite{Lee2010}. The PL spectrum in Fig.\ref{fig:optics-eq}b has three peaks at$\sim$1.65, 1.88 and 2.03eV. The position and energy difference between the two higher energy peaks match the A and B peaks measured with static absorbance in Fig.\ref{fig:Simulation-eq}a. The lower energy PL peak is attributed to emission from defects\cite{Chow2015}.

The static absorbance in Fig.\ref{fig:Simulation-eq}a identifies several excitonic peaks. A (1.9eV) and B (2.05eV) are associated to vertical transitions at K, from VB top to CB bottom\cite{Mak2010}. The 150meV energy separation is due to the spin-orbit interaction of K VB states\cite{Mak2010}. At higher energies, there is a broad contribution due to the C exciton at 2.9eV\cite{Kozawa2014,Qiu2013,Dumcenco2015}. The large (hundreds meV) broadening of the C peak originates from multiple e-h optical transitions in the region of the Brillouin zone (BZ) where VB and CB have the same dispersion\cite{Qiu2013,Carvalho2013}. The asymmetry of the peak is due to an additional contribution of the D excitonic peak centered at 3.1eV\cite{Li2014}, with a lower oscillator strength than the C exciton. This peak was assigned to Van Hove singularities\cite{Dumcenco2015} or to impurities\cite{Vella2016,Camellini2018}. The absorption in Fig.\ref{fig:Simulation-eq}a is well fitted by a sum of four Lorentz oscillators (one for each exciton peak).

Absorbance and band structure of 1L-MoS$_2$ are computed by ab-initio many-body perturbation theory, solving the Bethe-Salpeter Equation\cite{Onida2002} on top of a density functional theory (DFT) calculation and including quasi-particle corrections\cite{Onida2002}. A full spinorial approach is used to take into account the strong spin-orbit coupling\cite{Sangalli2019} (See Methods). The energies and relative intensities of the calculated excitonic peaks are in agreement with experiments, as shown in Fig.\ref{fig:Simulation-eq}a.

The C exciton is analyzed in Fig.\ref{fig:Simulation-eq}b. It involves a broad spectrum of transitions in the band-nesting region between K and $\Gamma$ and in the region between $\Gamma$ and M\cite{Molina-Sanchez2013,Qiu2013}. In Figs.\ref{fig:Simulation-eq}c-d the modulus of the excitonic wavefunction in reciprocal space is plotted on the quasi-particle band structure and in the BZ.

We then photo-excite the sample with frequency-tunable ultrashort pulses$<$20fs, shorter than the inverse of the frequency of the $A'_{1}$ and $E'$ optical phonon modes of 1L-MoS$_2$\cite{Molina-Sanchez2011}. The differential transmission ($\Delta T/T$) is measured by a white-light supercontinuum probe pulse whose spectral content covers all excitonic resonances. The fluence is such that the maximum amplitude of $\Delta T/T$ is$<$1$\%$. Since the transmission in 1L-TMDs dominates over the reflectivity close to the excitonic resonances\cite{Li2014}, the $\Delta T/T$ and TA spectra $\Delta A$ are related according to: $\Delta A=-ln(1+\Delta T/T)\sim-\Delta T/T$\cite{Soavi2013}.
\begin{figure*}
\centerline{\includegraphics[width=180mm]{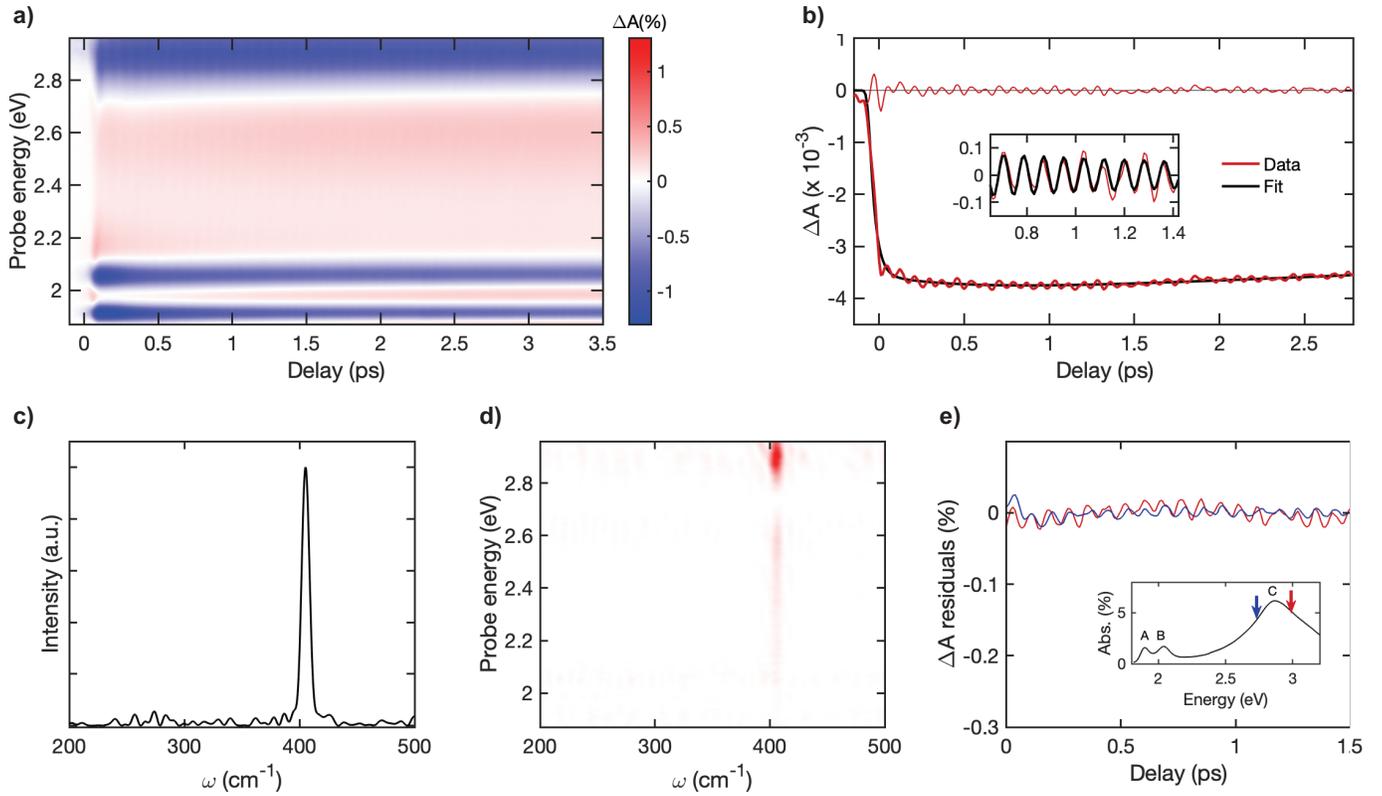}}
\caption{a) $\Delta A(\omega,\tau)$ map as a function of probe energy and pump-probe delay time. b) (red) Temporal cut of $\Delta A(\omega,\tau)$ map at a probe energy of 2.9eV, corresponding to the maximum of the C exciton PB signal. (Black) Fit of experimental data. The oscillating part (residuals) is obtained by subtracting the fit from the data. c) FT of the residual displaying the $A'_{1}$ mode. d) FT of the coherent part of the signal at different probe energies. The $A'_{1}$ peak has a maximum around the C exciton optical transition. e) Residuals at probe energies lower and higher than the C peak, as shown in the inset.}
\label{fig:transient_absorption_exp}
\end{figure*}
\begin{figure*}
\centerline{\includegraphics[width=180mm]{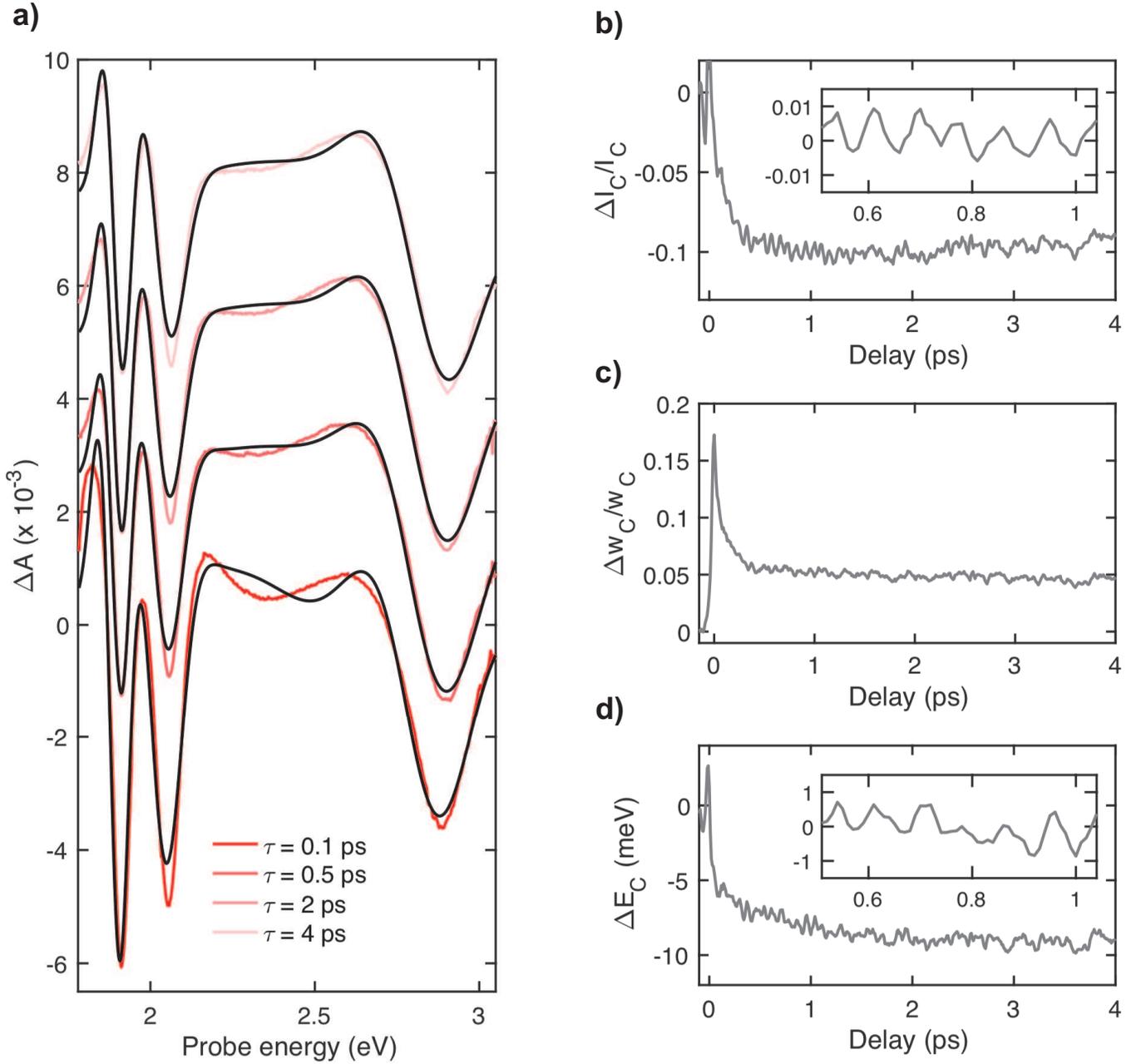}}
\caption{a) TA spectra at different delay times. The black lines are the differential fits obtained by modification of the fitting parameters of the static absorption in Fig.\ref{fig:optics-eq}. b,c,d) Temporal evolution of the fitting parameters of the C exciton peak normalized to the equilibrium values. The insets show that both oscillator strength I$_c$ and energy E$_c$ of the C peak are coherently modulated with a period matching the $A'_{1}$ phonon}
\label{Fig:FigFit}
\end{figure*}
\begin{figure*}
\includegraphics[width=160mm]{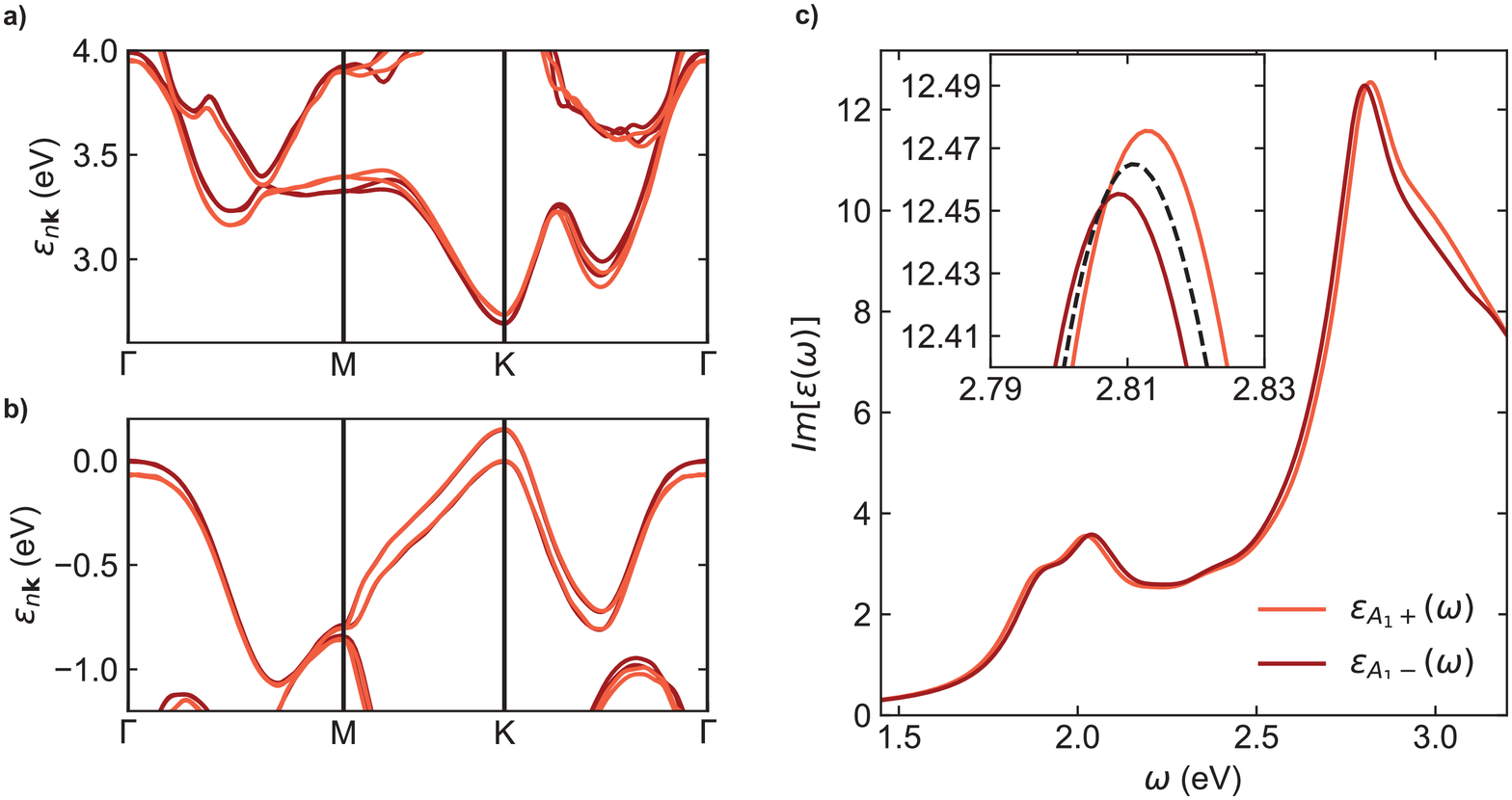}
\caption{Changes in (a,b) band-structure, and (c) in absorbance $A(\omega)$, due to atoms displacement along the $A_{1g}$ phonon mode for two opposite directions. The changes are enhanced by a factor 5. In the inset of (c), the change of the absorption peak compared to the equilibrium position (dashed line) is shown without enhancement.}
\label{fig:optics-displaced}
\end{figure*}

Fig.\ref{fig:transient_absorption_exp}a plots the $\Delta A(\omega,\tau)$ map at 300K as a function of probe energy and pump-probe delay. As previously reported\cite{Pogna2016,Cunningham2017}, $\Delta A$ exhibits a series of negative peaks, centered around the energy of the three exciton peaks, ascribed to the photobleaching (PB) of the excitonic resonances. For each negative peak, $\Delta A$ has the same number of positive features, referred to as photo-induced absorption, redshifted with respect to the PB peaks\cite{Pogna2016}. The derivative shape of $\Delta A(\omega)$ is the result of many-body effects enhanced by strong Coulomb interaction, inducing an energy shift and a modification of the shape of each excitonic peak. Photo-excitation of carriers increases Coulomb screening and modifies the exchange-correlation effects, leading to a renormalization of quasi-particle bandgap and exciton binding energy\cite{Pogna2016}. These two effects give rise to opposite shifts of the excitonic peaks which partially compensate\cite{Pogna2016} resulting in an overall redshift of all excitons. Phase space filling, caused by Pauli blocking of photoexcited excitons, also reduces the exciton oscillator strength, inducing PB. A transient increase of the spectral broadening $\gamma$ of the excitonic peak also contributes to the transient optical response\cite{Ruppert2017}. As shown in Fig.\ref{fig:transient_absorption_exp}c, the optical response is periodically modulated, during the first few ps by fast (i.e with sub-100-fs period) oscillations, particularly pronounced at 2.9eV (i.e. the maximum of the C exciton PB signal), and strongly decreasing in amplitude when approaching the lower energy peaks.

We now focus on the oscillatory component of the optical response of 1L-MoS$_2$ in an energy range close to the C exciton. For each probe photon energy, we fit the $\Delta A(\tau)$ dynamics with a convolution of instrumental response function and sum of an exponential build up and an exponential decay dynamics. In this way, the oscillatory component of the signal is isolated by subtracting the slowly varying background from the experimental data, Fig.\ref{fig:transient_absorption_exp}b. The frequency of the residual oscillations is determined by performing a Fourier Transform (FT) analysis. The FT spectrum in Fig.\ref{fig:transient_absorption_exp}c reveals a sharp peak$\sim$406cm$^{-1}$, corresponding to an oscillation period$\sim$82fs, with a FWHM$\sim$6cm$^{-1}$. The frequency of this peak matches the higher energy Raman mode in Fig.\ref{fig:optics-eq}a. Therefore we assign it to the Raman active $A'_{1}$. Although the Raman $A'_{1}$ and $E'$ peaks have similar intensities in Fig.\ref{fig:optics-eq}a (their ratio is$\sim$1.5), the FT spectrum does not show significant contribution from $E'$. This is consistent with Ref.\citenum{Jeong2016}, which showed that the optical response around the A exciton is modulated only by the $A'_{1}$ phonon in 1L-WSe$_2$. The phonon dephasing time $\tau_{deph}$=1.7$\pm$0.2ps is estimated by fitting the temporal oscillations with a single exponentially decaying harmonic oscillator $A_{0}$exp(-t/$\tau_{deph}$)cos($\Omega$+$\varphi$) where $A_{0}$, $\Omega$ and $\varphi$ are, respectively, amplitude, frequency and phase of modulation\cite{Jeong2016}. $\tau_{deph}$ is almost three times smaller than that measured at the same temperature for CP in exfoliated 1L-WSe$_2$\cite{Jeong2016}. This can be explained in terms of different defect concentration in exfoliated and CVD-grown 1L-TMDs. In the former case, the CP dephasing process is mainly determined by anharmonic decay, due to phonon-phonon interactions\cite{Jeong2016}, while in the latter, an additional dephasing due to scattering between phonons and defects is active\cite{Hase2000}.

The maximum CP oscillation amplitude is$\sim3\times10^{-4}$ (i.e.$\sim$2$\%$ of the peak value of the PB signal). This is almost two orders of magnitude higher than the amplitude of optical phonon oscillations reported previously in 1L-TMDs\cite{Jeong2016}. To get an insight into the CP detection mechanism, we exploit the broad bandwidth of the probe pulse and we study the probe energy dependence of the oscillatory component of the TA signal. Fig.\ref{fig:transient_absorption_exp}d plots the FT of the residual as a function of probe energy. The amplitude of the oscillations strongly depends on the probe energy, with a maximum around the C exciton. Across the A and B excitons, no appreciable oscillatory signal is detected, probably because the amplitude of such oscillations lies below the sensitivity of our apparatus ($\sim$10$^{-5}$). On the other hand, the phase of the coherent oscillations does not vary across the $C$ excitonic transition. Fig.\ref{fig:transient_absorption_exp}e reports the residuals of the fit for probe energies on the red and blue edges of the C exciton. Both traces shows almost in-phase oscillations.

The study of the temporal evolution of the excitonic line shape can provide additional insights into the CP physics. The pump-induced change of the optical absorption is the result of different contributions, i.e. the transient change of energy $E$, width $w$ and intensity $I$ of each exciton. In order to disentangle these contributions, we perform a differential fit of the $\Delta A$ spectrum at each delay time. This reproduces the shape of the $\Delta A$ spectrum in Fig.\ref{Fig:FigFit}a. We focus on the transient optical response around the C peak. The temporal dynamics of the fitting parameters of the C exciton are in Fig.\ref{Fig:FigFit}b-d. The parameters are normalized to the values obtained by fitting the static absorption, as shown in Fig.\ref{fig:Simulation-eq}a. After photoexcitation, we observe a transient reduction of the intensity of the excitonic transition, a redshift and an increase of line-width. The temporal dynamics (build-up and relaxation to equilibrium) of intensity and energy fitting parameters are coherently modulated. A modulation of the exciton transition energy around the equilibrium position would result in a $\pi$ phase shift of the oscillations probed at the red and blue edges of the C peak. Conversely, an intensity modulation of the peak results in an in-phase modulation through the scanned probe energy range. Since no flip of the phase sign of the oscillations is detected over the probed spectral window (Fig.\ref{fig:transient_absorption_exp}e), intensity modulation dominates over energy modulation.

In order to understand the origin of the coherent oscillations and their enhancement around the C exciton, we quantify, using \textit{ab initio} calculations, the impact of atomic displacement on band structure and absorption of 1L-MoS$_2$. Following Ref.\citenum{Zeiger1992}, $\Delta A(\omega,\tau)$ can be factorized into two contributions. The first is electronic, $\Delta A[f_{nk}(\tau)](\omega)$, with temporal evolution non oscillatory and dependent on the non-equilibrium electronic populations $f_{nk}(\tau)$ created by the pump pulse. The second, $\Delta A[R(\tau)](\omega)$, describes the effect of the collective atomic motion initiated by the ultrashort pump pulse and has an oscillatory behavior. $R(\tau)$ describes the time evolution of the atoms displaced out of equilibrium after the action of the pump. In Ref.\citenum{Pogna2016} we showed that $\Delta A[f_{nk}(\tau)](\omega)$ is dominated by band gap renormalization and Pauli blocking of the excitonic peaks. Now we focus on the simulation of the oscillatory part of the transient optical response $\Delta A[R(\tau)](\omega)$. We assume that the atoms oscillate according to $R(\tau)\approx R_0+(R_X-R_0)cos(\omega_X\tau)$, where $R_X$ is the atomic position when they are displaced along the $A'_{1}$ phonon eigenvector $u_{A'_{1}}$, with $R_{0}$ the nuclear coordinate before pump excitation.

The band structure and optical absorption are computed at the two extrema of these oscillations. We refer to the two displaced positions as $R_{A'_1}^{\pm}$. In our calculation, all the DFT wave-functions and energies, quasi-particle corrections and excitonic peaks are computed in the lattice displaced configuration\cite{Miranda2017}. The results are in Fig.\ref{fig:optics-displaced}. The VB and CB nearby K are almost unaffected by atomic displacement. On the other hand, the electronic states in the band nesting region between $\Gamma$-M and $\Gamma$-K, which contribute to the C exciton, are strongly influenced by the $A'_{1}$ phonon displacement. As a consequence, the A and B excitonic peaks are weakly affected by the lattice displacement, while the C exciton is strongly renormalized. The main effect is an energy shift of the C exciton. The overall shift emerges from an interplay between two effects which partially cancel each other: changes in the quasi-particle band-structure and variation of exciton binding energy. The former prevails. Together with a shift of the C peak, a change in the intensity is also observed in the inset of Fig.\ref{fig:optics-displaced}c.
\begin{figure}
\includegraphics[width=85mm]{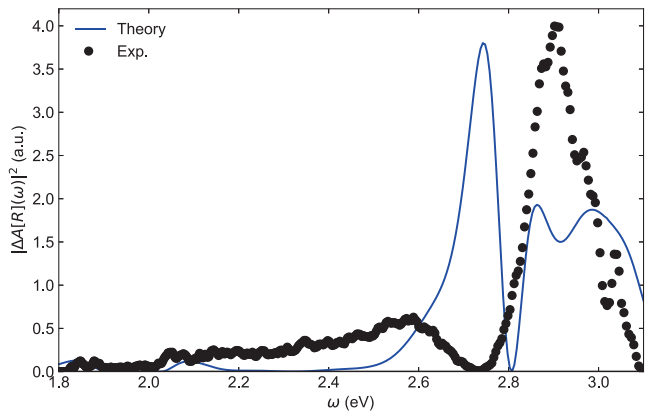}
\caption{Probe photon energy dependence of the amplitudes of the $A'_{1}$ coherent oscillations (black dots) compared with the square modulus of the variation in the absorbance due to atoms displacement}
\label{fig:SpectralDependence}
\end{figure}

$\Delta A[R(\tau)](\omega)$ can be reconstructed by computing the difference between the absorption spectrum calculated for two displaced atom configurations. As shown in Fig.\ref{fig:SpectralDependence}, the energy profile of $\Delta A[R(\tau)](\omega)$ reproduces the spectral dependence of the CP amplitude (vertical cut of Fig.\ref{fig:transient_absorption_exp}d at the $A'_{1}$ frequency), confirming a strong enhancement of the oscillations around the C exciton. The simulations do not fully reproduce the probe photon energy dependence of the CP phase. While the fit of the C exciton lineshape indicates that intensity oscillation is predominant with respect to energy modulation, the simulations point towards an energy modulation scenario resulting in a phase inversion of the oscillations across the C exciton (the node of the calculated curve in Fig.\ref{fig:SpectralDependence} is due to a $\pi$ phase flip of the oscillations around the C peak). We attribute this to an overestimation of the C peak energy shift caused by a not exact cancelation of the two opposite energy shifts discussed above.

We focus now on the generation of CP in 1L-MoS$_2$. Two mechanisms were proposed: Impulsive Stimulated Raman Scattering (ISRS)\cite{Dhar1994} and Displacive Excitation of Coherent Phonons (DECP)\cite{Cheng1991,Zeiger1992}. ISRS was used to describe CP excitation in the transparency regime\cite{Dhar1994}. In ISRS, all Raman active phonon modes are impulsively excited, provided that the spectral bandwidth of the laser pulse exceeds the frequency of the phonon mode. DECP was used to explain why, in the absorption regime, only fully symmetric modes (i.e.$A'_{1}$) can be coherently excited, although different symmetry modes (i.e. $E'$) have comparable intensity in the CW Raman spectrum. The absence of the lower symmetry $E'$ modes in the transient optical response and the different initial phase of the oscillations was interpreted as evidence that DECP excitation cannot be reduced to a Raman scattering process\cite{Zeiger1992}. Refs.\citenum{Garrett1996,Merlin1997,Stevens2002} tried to unify the two mechanisms, and proposed a new approach treating the DECP model as a particular case of the ISRS\cite{Garrett1996,Stevens2002}. Within this approach, referred as transient stimulated Raman scattering (TRSR)\cite{Stevens2002}, CP generation is controlled by two different tensors characterized by the same real parts, but different imaginary parts. The first tensor describes the coupling between phonons and virtual electronic transitions, and gives rise to an impulsive driving force, while the second accounts for the coupling between phonons and real charge density fluctuations, and results in a displacive driving force\cite{Stevens2002}. Up to now, there is no clear evidence of which is the correct formalism to describe CP excitation in the absorptive regime\cite{Riffe2007,Misochko2004,Lobad2001}. The CP phase is not a good parameter to discriminate between DECP and TRSR, because they both give the same prediction. In our case, the uncertainty in the determination of the zero delay does not allow to precisely estimate the CP phase. A $\frac{\pi}{4}$ phase shift, the phase difference between cosine and sine oscillations, corresponds to a temporal uncertainty$\sim$20fs (comparable to the temporal resolution of our experiment). Our approach is then to measure the dependence of the $A'_{1}$ CP amplitude on pump energy and correlate it with the corresponding Raman cross section.
\begin{figure*}
\includegraphics[width=180mm]{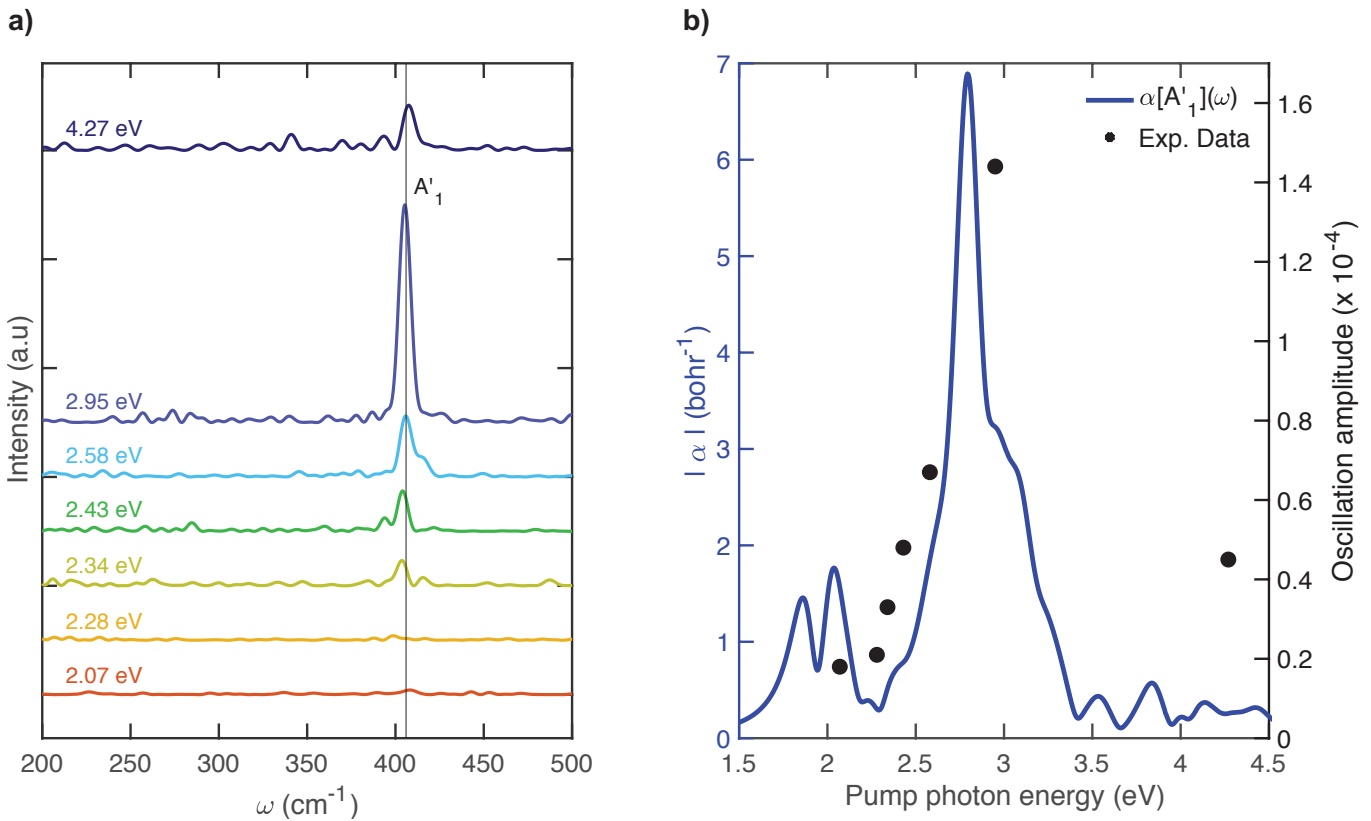}
\caption{a) FT spectrum of the temporal cut for 2.9eV probe energy at different pump energies. b) $\alpha(\omega)$ for $A'_1$ phonon (blue), compared with the amplitude of the coherent oscillations at different pump photon energies (black dots)}
\label{fig:ft-raman}
\end{figure*}

The intensity of the FT spectrum of the oscillating part of the PB signal at 2.9eV is reported as a function of pump energy in Fig.\ref{fig:ft-raman}. The pump pulse photon energy is tuned across a broad range between 2 and 4.3eV, without changing its temporal duration, while the incident fluence is adjusted to maintain the same maximum variation of the non-oscillatory signal. As explained in Methods, we use different broadband nonlinear frequency conversion processes, coupled to dispersion compensation, to generate frequency-tunable sub-20fs pump pulses in the vis-UV range. As for the detection process, the data reveal a resonant behavior and show that the amplitude of the $A'_{1}$ oscillations has a maximum at the C exciton, decreasing at higher energy. The continuous line in Fig.\ref{fig:ft-raman}b is the calculated amplitude of the Raman tensor $\alpha(\omega)$ for the $A'_{1}$ mode as a function of energy. $\alpha(\omega)$ is the change of the dielectric susceptibility $\chi$ with respect to the nuclear coordinates ($\alpha=\partial\chi/\partial R$) and it is computed in the framework of finite-differences of the dielectric function\cite{Gillet2013}, also including excitonic effects\cite{Miranda2017}. Similar to the CP amplitude, $\alpha$ has a resonant profile peaked around the C exciton. The lower energy part of $\alpha(\omega)$, around the B exciton, deviates from the experimental data. This is due to the fact that the calculation of $\alpha$ based on the finite difference method tends to overestimate the Raman response around the A and B excitons\cite{Crespi2018}. The spectral dependence of $\alpha(\omega)$ is confirmed by resonant Raman spectroscopy measurements showing a strong enhancement of both optical phonon modes when the excitation laser is on resonance with the C exciton\cite{Carvalho2015,Carvalho2016,Scheuschner2015,Lee2015}. The comparison between the spectral dependence of the CP amplitude and the spontaneous Raman tensor, sheds new light on the nature of CP generation in 1L-TMDs.

Thus, CP generation in 1L-TMDs shares typical features of both DECP and TRSR mechanisms. On one hand, the absence of the lower symmetry mode $E'$ in the FT spectrum is in contrast with the ratio between the intensities of $A'_{1}$ and $E'$ peaks in Raman spectroscopy and can be explained within DECP. On the other hand, the CP amplitude and calculated Raman tensor display the same resonant profile on excitation energy around the C exciton. The same resonant behavior is reported also for the detection of $A'_{1}$ CP, meaning that both detection and excitation of $A'_{1}$ CP are governed by the same Raman-like mechanism.

In summary, we studied CP generation and detection in 1L-MoS$_2$ by high time resolution broadband pump-probe spectroscopy. We demonstrated that $A'_{1}$ optical phonon modes strongly couple with the C exciton, giving rise to a temporal modulation of the TA response over a broad energy range around this excitonic peak. The excitation profile of such coherent oscillations follows the energy dependence of the Raman tensor. This proves that the lattice coherence in 1L-MoS$_2$ can be described as a Raman-like excitation mechanism. Ab-initio calculations of the band structure for displaced atoms configuration confirm the strong coupling between C exciton and $A'_{1}$ phonons. Our results pave the way for the coherent control of optical phonons in layered materials on a sub picosecond time scale and they enable the study of exciton-phonon interaction on the atomic scale.

We acknowledge funding from National Research Fund, Luxembourg (Projects C14/MS/773152/ FAST-2DMAT and INTER/ANR/13/20/NANOTMD), the Juan de la Cierva Program from MINECO-Spain (Grant IJCI-2015-25799), the EU Graphene and Quantum Flagships, ERC grant Hetero2D, EPSRC grants EP/L016087/1, EP/K01711X/1, EP/K017144/1, the Early Postdoc Mobility program of the Swiss National Science Foundation (P2BSP2 168747), the EU project MaX Materials design at the eXascale H2020-EINFRA-2015-1, grant agreement no. 676598, and Nanoscience Foundries and Fine Analysis - Europe H2020-INFRAIA-2014-2015, grant agreement no. 654360.
\section{Methods}
\subsection{Ultrafast Pump-Probe Spectroscopy}
The laser source for the pump-probe experiments is an amplified Ti:Sapphire (Coherent Libra II), which emits 100fs pulses at 1.55eV. The repetition rate is 2kHz and the average output power is 4W. A fraction of the total power (300mW) is used for the experiment. Tunable pump pulses with photon energy ranging from 2 to 4.3eV are generated by a Non-collinear Optical Parametric Amplifier (NOPA)\cite{Manzoni2016}. The pump pulses are temporally compressed by 2 chirped mirrors in the visible and by a MgF$_2$ prism compressor in the UV\cite{Borrego2018}. The pulse characterization is performed by frequency-resolved optical gating in the visible, and by a two-dimensional spectral-shearing interferometry method (2DSI) in the UV\cite{Borrego2015}. For all pump photon energies, the temporal duration is$<$20fs. This sets the temporal resolution of the pump-probe setup\cite{Polli2010}. The pump fluence is kept$\sim$10$\mu J/cm^2$, leading to a photogenerated carrier density$\sim$6-9$\times$10$^{11}$cm$^{-2}$, depending on pump photon energy. The pump beam is modulated at 1kHz by a mechanical chopper. The overall detection sensitivity is$\sim$10$^{-5}$, with a 2s integration time. The broadband white light probe, characterized by a spectral content ranging between 1.8 and 3.0eV, is generated by focusing 1.55eV pulses on a 3mm CaF$_2$ plate. The fundamental frequency of the laser is filtered by a colored glass short-pass filter. The fluctuations of the laser are$<$0.2$\%$. In all experiments, pump and probe beams have parallel linear polarizations. The two pulses are temporally synchronized by a motorized translation stage. The spot sizes of pump and probe beams on the sample are respectively$\sim$150$\mu m$ and$\sim$100$\mu m$. The transmitted probe light is collected with a 75mm lens, dispersed by a CaF$_2$ prism and detected by a Si spectrometer working at the full laser repetition rate. All the TA maps are corrected for the chirp introduced by white light generation.
\subsection{Ab-initio simulations}
To compute the band structure, the optical properties and their variation we perform first-principles DFT simulations within the local density approximation using Quantum Espresso\cite{Giannozzi2009} and Yambo\cite{Marini2009,Sangalli2019}. Ground state self-consistent (SCF) simulations are done with {{\normalfont\ttfamily pw.x}}\cite{Giannozzi2009} to get the converged density of 1L-MoS$_2$ and the equilibrium atomic positions. We use pseudo-potentials with 6-active electrons for Mo and 6 from S, an energy cut-off of $80\ Ry$ and $16\times16\times1$ k-points mesh. Then we compute the phonons energies and eigevectors with {{\normalfont\ttfamily ph.x}}\cite{Giannozzi2009} at q=0 on the same grid. A subsequent non-self consistent (NSCF) calculation on a $36\times 36\times 1$ k-points grid is done to generate the wave-functions and the energies to be used by {{\normalfont\ttfamily yambo}}\cite{Marini2009} and another along a BZ path to plot the band-structure. Quasi-particle energies are computed within the GW approximation for the self-energy\cite{Onida2002}. The plasmon-pole approximation is used for the dynamically screened interaction. The absorption spectrum is computed solving the Bethe-Salpeter equation\cite{Onida2002}, including local fields and excitonic effects, considering all transitions with energy$\leq$5eV, and a cutoff of 2Ry for the local-fields, and screened interaction accounting for the exciton binding energy. A Coulomb cut-off technique\cite{Sangalli2019} is used in all calculation steps. To compute the variations induced by the coherent oscillations the atoms are displaced from their equilibrium position by 0.01 Bohr along the phonon eigenmode (i.e. $R_X-R_0= 0.01 u_{A_{1g}}$ Bohr). This corresponds to a displacement of S$\sim7\times 10^{-3}$ Bohr, guaranteeing that we are in a linear regime with respect to the perturbation, but still large enough to avoid numerical noise. We also verify that $2|A[R_{A_{1g}}^+](\omega)-A[R_0](\omega)|=2|A[R_{A_{1g}}^-](\omega)-A[R_0](\omega)|=|[A[R_{A_{1g}}^+](\omega)-A[R_{A_{1g}}^-](\omega)]|$. All the steps, SCF, NSCF calculations, quasi-particles corrections and BSE solution are repeated with the same parameters used for the equilibrium geometry. The flow of simulations is handled via the yambopy utility\cite{Sangalli2019}.

\end{document}